
\magnification=\magstep1
\hsize=6.75truein
\vsize=8.7 truein
\baselineskip=16pt
\parskip=\medskipamount
\parindent= .35in
\def \di{\partial}

\def \gA {{\bf g}_{A}}
\def \gY {{\bf g}_{Y}}
\def \gYA {{\bf g}_{A}^Y}
\def \gAY {{\bf g}_{Y}^A}
\def \a {\alpha}
\def \b {\beta}
\def \d {\delta}
\def \g {\gamma}

\def \l {\lambda}
\def \o {\omega}
\def \s  {\sigma}
\def \t  {\tau}
\def \tr{{\rm tr}}
\def \Stab{{\rm Stab}}
\def \det{{\rm det}}
\def \Ad{{\rm Ad}}
\def \log {{\rm log}}
\def \red {{\rm red}}
\def \diag{{\rm diag}}
\def \ra {\rightarrow}
\def \lra{\longrightarrow}

\def \lmt {\longmapsto}
\def \wt {\widetilde}
\def \ss{\subset}
\def \IYA {{\cal I}_A^Y}
\def \IAY {{\cal I}_Y^A}
\def \scst {\scriptstyle}

\font\eightrm=cmr8
\font\eightbf=cmbx8
\font\eightit=cmti8
\font\grand=cmbx10 at 14.4truept
\def \smalltype{\let\rm=\eightrm  \let\bf=\eightbf
\let\it=\eightit \let\sl=\eightsl \let\mus=\eightmus
\baselineskip = 9.5pt minus .75pt  \rm}
\rightline{ CRM-1844 (1993) \break}
\bigskip
\centerline {\grand  Dual Isomonodromic Deformations}
\centerline{{\grand and Moment Maps to Loop Algebras}
\footnote{$^\dagger$}{\smalltype
Research supported in part by the Natural Sciences and Engineering Research
Council of
Canada.}}
 \bigskip \centerline{ J. Harnad\footnote{$^\ddagger$}{\smalltype e-mail
address:
harnad@alcor.concordia.ca {\it \ or\ } harnad@mathcn.umontreal.ca}}
\smallskip
\centerline
{\smalltype \it Department of Mathematics and Statistics, Concordia
University}
\centerline{\smalltype \it  7141 Sherbrooke W., Montr\'eal, Canada H4B 1R6, \
and}
\centerline{\smalltype \it  Centre de recherches math\'ematiques,
Universit\'e de Montr\'eal}
\centerline{\smalltype \it  C.~P. 6128-A, Montr\'eal, Canada
H3C 3J7}
  \bigskip
{\smalltype \centerline{\bf Abstract}
\smallskip
The Hamiltonian structure of the monodromy preserving deformation equations
of Jimbo {\it et al }  {\bf [JMMS]} is explained in terms of parameter
dependent pairs of moment maps from a symplectic vector space to the dual
spaces of two different loop algebras. The nonautonomous Hamiltonian systems
generating the deformations are obtained by pulling back spectral invariants
on Poisson subspaces consisting of elements that are rational in the loop
parameter and identifying the deformation parameters with those determining
the moment maps. This construction is shown to lead to ``dual'' pairs
of matrix differential operators whose monodromy is preserved under the same
family of deformations. As illustrative examples, involving discrete and
continuous reductions, a higher rank generalization of the Hamiltonian
equations governing the correlation functions for an impenetrable Bose gas
is obtained, as well as dual pairs of isomonodromy representations for the
equations of the Painleve transcendents $P_{V}$ and $P_{VI}$.}
\baselineskip 14pt
\bigskip
\noindent {\bf 1. Monodromy Preserving Hamiltonian Systems} \hfill

   The following integrable Pfaffian system was studied by Jimbo, Miwa,
M\^ori and Sato in {\bf [JMMS]}:
$$
dN_i = -\sum_{j=1 \atop j \ne i}^n [N_i, N_j] d\log(\a_i - \a_j)
    -[N_i, d(\a_i Y) + \Theta].  \eqno(1.1)
$$
 Here $\{N_i(\a_1, \dots,\a_n,\ y_1, \dots, y_r)\}_{i=1, \dots n}$
is a set of $r\times r$ matrix functions of $n+r$ (real or complex) variables
$\{\a_i, y_a\}_{i=1, \dots n \atop a=1, \dots r}$, $Y$ is the diagonal $r
\times r$
matrix $Y=\diag(y_1, \dots, y_r)$ and the matrix differential form $\Theta$ is
defined by:
$$
\Theta_{ab}=(1-\d_{ab})(\sum_{i=1}^nN_i)_{ab} d\log(y_a -y_b).  \eqno(1.2)
$$
This system determines deformations of the differential operator:
$$
{\cal D}_{\l} := {\di \over \di \l} - {\cal N}(\l),  \eqno (1.3)
$$
where
$$
{\cal N}(\l) := Y + \sum_{i=1}^n{N_i \over \l- \a_i},  \eqno(1.4)
$$
that preserve its monodromy around the regular singular points $\{\l
=\a_i\}_{i=1,\dots
n}$ and at $\l=\infty$.

  It was observed in {\bf [JMMS]} (appendix 5) that, expressing the $N_i$'s as
$$
N_i = G_i^TF_i,  \eqno(1.5)
$$
where $\{F_i, G_i\in M^{k_i \times r}\}_{i=1,\dots n}$ are pairs of maximal
rank
$k_i\times r$ rectangular matrices ($k_i \leq r$), with $\{F_iG_i^T = L_i \in
gl(k_i)\}_{i=1, \dots n}$ constant matrices related to the monodromy of ${\cal
D}_\l$ at
$\{\a_i\}_{i=1,\dots n}$, eq.~(1.1) may be expressed as a set of compatible
nonautonomous
Hamiltonian systems:
$$
\eqalignno
{dF_i&=\{F_i, \omega\} & (1.6a) \cr
dG_i &= \{G_i, \omega\}.& (1.6b)   \cr}
$$
Here $d$ denotes the total differential with respect to the variables
$\{\a_i, y_a\}_{i=1, \dots n \atop a=1, \dots r}$, the $1-$form $\omega$ is
defined
as:
$$
\omega := \sum_{i=1}^nH_id\a_i + \sum_{a=1}^r K_a dy_a,  \eqno(1.7)
$$
with
$$
\eqalignno{
H_i &:= \tr(YN_i) + \sum_{j=1 \atop j\ne i}^n {\tr(N_iN_j) \over \a_i -\a_j}
  ,\qquad i=1, \dots, n &(1.8a)   \cr
 K_a &:= \sum_{i=1}^n \a_i (N_i)_{aa} +
\sum_{b=1 \atop b \ne a}^r {(\sum_{i=1}^n N_i)_{ab} (\sum_{j=1}^n N_j)_{ba}
\over y_a -
y_b}, \quad  a=1, \dots, r &(1.8b).}
$$
and the Poisson brackets in the space of $(F_i, G_i)$'s are defined to be such
that the
matrix elements of $\{F_i, G_i\}_{i=1, \dots n}$ are canonically conjugate:
$$
 \eqalignno{
\{(F_i)_{a_i a}, (G_j)_{b_i b}\} &= \d_{ij} \d_{a_i b_j} \d_{ab}  &(1.9)  \cr
 i,j =1, \dots,n,\quad a,b =1, &\dots ,r, \quad  a_i, b_i=1, \dots, k_i. }
$$
The Frobenius integrability of the differential system (1.1) follows from the
fact that
the Hamiltonians $\{H_i, K_a\}_{i=1, \dots n \atop a=1, \dots r}$ Poisson
commute.
It was also noted in {\bf [JMMS]} that, with respect to the Poisson brackets
(1.9), the
matrices $\{N_i\}_{i=1, \dots, n}$ satisfy:
$$
\{(N_i)_{ab},(N_j)_{cd}\} =\d_{ij}\left(\d_{bc}(N_i)_{ad} - \d_{ad}
(N_i)_{cb}\right),
\eqno(1.10)
$$
which is the Lie Poisson bracket on the space $(\oplus_{i=1}^n gl(r))^*$ dual
to the
direct sum of $n$ copies of the Lie algebra $gl(r)$ with itself. The
$1$-form $\omega$ is exact on the parameter space and may be
interpreted as the logarithmic differential of the $\tau$-function,
$$
\omega = d\log \tau.  \eqno(1.11)
$$

     Numerous applications of such systems exist; in particular, in the
calculation of
correlation functions for integrable models in statistical mechanics and
quantum field
theory {\bf [JMMS, IIKS]}, in matrix models of two dimensional quantum  gravity
{\bf [M]} and
in the computation of level spacing distribution functions for random matrix
ensembles {\bf
[TW1, TW2]}.

    In the next section these systems will be examined within the context of
loop algebras, using
an approach originally developed for the autonomous case, involving isospectral
flows, in {\bf [AHP,
AHH1]}. This is based on ``dual'' pairs of parameter dependent  moment maps
from symplectic vector
spaces to two different loop algebras $\wt{gl}(r)$ and $\wt{gl}(N)$, where
$N=\sum_{i=1}^n k_i$. The nonautonomous Hamiltonian systems (1.6)--(1.8) will
be generated by
pulling back certain spectral invariants, viewed as polynomial functions on
rational coadjoint
orbits, under these moment maps,  and identifying the parameters determining
the maps with the
deformation parameters of the system. This construction leads to a pair of
``dual'' first order
matrix differential operators with regular singular points at finite values of
the spectral
parameter, both of whose monodromy data are invariant under the deformations
generated by these
Hamiltonian systems. In section 3, the generic systems so obtained will be
reduced under various
discrete and continuous Hamiltonian symmetry groups. A rank $r=2s$
generalization of the
systems determining the correlation functions for an impenetrable Bose gas (or,
equivalently, the generating function for the level spacing distribution
functions for
random matrix ensembles {\bf [TW1]}) will be derived by reduction to the
symplectic loop
algebra $\wt{sp}(2s)$. The ``dual'' isomonodromy representations of the
equations for the
Painlev\'e transcendents $P_{V}$ and $P_{VI}$ will also be derived, and their
Hamiltonian
structure deduced through  reductions under continuous groups. A brief
discussion of
generalizations to systems with irregular singular points is given in section
4.

\bigskip \noindent {\bf 2.  Loop Algebra Moment Maps and Spectral Invariants}
\hfill

  In {\bf [AHP, AHH1]}, an approach to the embedding of finite
dimensional integrable systems into rational coadjoint orbits of loop algebras
was
developed, based on a parametric family of equivariant moment maps $\wt{J}_A :M
\lra
\wt{gl}(r)_{+}^*$ from the space  $M= \{F,G \in M^{N \times r}\}$ of pairs of
$N \times
r$ rectangular matrices, with canonical symplectic structure
$$
\omega = \tr(dF \wedge dG^T) \eqno(2.1)
$$
to the dual of the positive half of the loop algebra $\wt{gl}(r)$. The maps
$\wt{J}_A$, which are parametrized by the choice of an $N\times N$ matrix $A\in
M^{N \times
N}$ with generalized eigenspaces of dimension $\{k_i\}_{i=1, \dots n}, \
\sum_{i=1}^n k_i =N$ and eigenvalues $\{\a_i\}_{i=1, \dots n}$, are defined by:
$$
\wt{J}_A :(F,G) \lmt G^T(A-\l I_N)^{-1}F  \eqno(2.2)
$$
where $I_N$ denotes the $N\times N$ identity matrix.
The conventions here are such that all the eigenvalues $\{\a_i\}_{i=1, \dots
n}$ are
interior to a circle $S^1$ in the complex $\l$-plane on which the loop algebra
elements
$X(\l) \in \wt{gl}(r)$ are defined. The two subalgebras $\wt{gl}(r)_{+},
\wt{gl}(r)_{-}$
consist of elements $X_+ \in  \wt{gl}(r)_{+}$, $X_- \in  \wt{gl}(r)_{-}$
that admit  holomorphic extensions, respectively, to the interior and exterior
regions,
with $X_- (\infty)=0$. The space $ \wt{gl}(r)$ is identified as a dense
subspace of its
dual space $\wt{gl}(r)^*$, through the pairing
$$ \eqalignno{
<X_1, X_2> &= \oint_{S^1} \tr\left(X_1(\l)X_2(\l)\right)d\l   & (2.3)\cr
X_1 &\in  \wt{gl}(r)^*, \ X_2 \in  \wt{gl}(r).}
$$
This also gives identifications of the dual spaces
$\wt{gl}(r)_{\pm}^*$ with the opposite subalgebras $\wt{gl}(r)_{\mp}$.

Taking the simplest case, when $A$ is diagonal, the image of the moment map is
$$
\eqalignno{
{\cal N}_0 (\l) &= G^T(A-\l I_N)^{-1}F= \sum_{i=1}^n {N_i \over \l - \a_i} , &
(2.4a) \cr
N_i &:= -G^T_i F_i,  &  (2.4b)}
$$
where $(F_i,G_i)$ are the $k_i\times r$ blocks in $(F,G)$ corresponding to the
eigenspaces
of $A$ with eigenvalues $\{\a_i\}_{i=1, \dots n}$. The set of all
${\cal N}_0 \in \wt{gl}(r)_{-}$ having the pole structure given in eq.~(2.4a)
forms a
Poisson subspace of $\wt{gl}(r)_-$, which we denote $\gA$. The coadjoint action
of the
loop group $\wt{Gl}(r)_+$ corresponding to the algebra $\wt{gl}(r)_{+}$,
restricted to
the subspace $\gA$, is given by:
$$
\eqalignno{
  Ad^*(\wt{Gl}(r)_{+}) :\gA &\lra \gA    \cr
Ad^*(g) :\sum_{i=1}^n {N_i \over \l - \a_i} &\lra
\sum_{i=1}^n {g_i N_i g_i^{-1} \over \l
-\a_i} & (2.5)  \cr g_i := g(\a_i),& \quad i=1, \dots , n.}
$$
We see that $\gA$ could equally have been identified with the dual space
$(\oplus_{j=1}^{n} gl(r))^*$ of the direct sum of $n$ copies of $gl(r)$ with
itself, and
the Lie Poisson bracket on $\wt{gl}(r)_{+}^* \sim \wt{gl}(r)_-$:
$$
\{f,\ g\}\vert_{{\cal N}_0} = <{\cal N}_0, [df, dg]\vert_{{\cal N}_0}>,
\eqno(2.6)
$$
reduces on the Poisson subspace $\gA$ to that for $(\oplus_{j=1}^{n} gl(r))^*$,
 as given
in eq.~(1.10).

    In the approach developed in {\bf [AHP, AHH1]}, one studies commuting
Hamiltonian flows on spaces of type $\gA$ (in general, rational Poisson
subspaces
involving higher order poles if the matrix $A$ is nondiagonalizable), generated
by
elements of the Poisson commuting spectral ring $\IYA$ of polynomials on
$\wt{gl}(r)^{*}$
invariant under the $\Ad^* \wt{Gl}(r)$--action (conjugation by loop group
elements),
restricted to the affine subspace $Y+ \gA$, where $Y\in gl(r)$ is some
fixed $r \times r$ matrix. The pullback of such Hamiltonians under $\wt{J}_A$
generates
commuting flows in $M$ that project to the quotient of $M$ by the Hamiltonian
action of
the stability subgroup $G_A:= Stab(A) \ss Gl(N)$. The Adler-Kostant-Symes (AKS)
theorem
then tells us that: \item{(i)} Any two elements of $\IYA$ Poisson commute (and
hence, so do
their pullbacks under the Poisson map $\wt{J}_A$).  \item{(ii)} Hamilton's
equations for
$H \in \IYA$ have the Lax pair form:
$$
 {d{\cal N} \over dt} = [(dH)_+, {\cal N}] = -
[(dH)_-, {\cal N}],   \eqno (2.7)
$$
 where
$$
{\cal N}(\l,t) := Y + {\cal N}_0(\l,t),  \eqno(2.8)
$$
with ${\cal N}_0 \in \wt{gl}(r)_-$ of the form (2.4a), $dH\vert_{\cal N}$
viewed as
an element of  $(\wt{gl}(r)^{*})^* \sim \wt{gl}(r)$, and the subscripts $\pm$
denoting
projections to the subspace $\wt{gl}(r)_{\pm}$.  \smallskip\noindent
The coefficients of the spectral curve of ${\cal N}(\l)$, determined by the
characteristic
equation
$$
\det ( Y + G^T(A-\l I_N)^{-1}  -z I_r) =0, \eqno(2.9)
$$
are the generators of the ring $\IYA$.

   In particular,  choosing
$$
Y := \diag(y_1, \dots , y_r)  \eqno(2.10)
$$
and defining $\{H_i \in \IYA\}_{i=1, \dots n}$ by:
$$
H_i({\cal N}) := {1\over 4 \pi i}\oint_{\l =\a_i}\tr(({\cal N}(\l))^2 d\l
=\tr(YN_i) +\sum_{j=1 \atop j\ne i}^n{\tr(N_iN_j)\over \a_i - \a_j}
   \eqno(2.11)
$$
(where $\oint_{\l =\a_i}$ denotes integration around a small loop containing
only this
pole), we see that these coincide with the $H_i$'s  defined in eq.~(1.8a).
Thus, the
Poisson commutativity of the $H_i$'s follows from the AKS theorem. The Lax form
of Hamilton's equations is:
$$
{\di{\cal N} \over \di t_i} = - [(dH_i)_-, {\cal N}]  \eqno(2.12)
$$
where
$$
(dH_i)_- = {N_i \over \l - \a_i} \in \wt{gl}(r)_-.  \eqno(2.13)
$$
Evaluating residues at each $\l=\a_i$, we see that this is equivalent to:
$$
\eqalignno{
{\di N_j\over \di t_i}&= {[N_j,N_i]\over \a_j - \a_i}, \quad j \neq i,
 \quad i,j = 1, \dots n.  &(2.14a)
 \cr
{\di N_i\over \di t_i} &= [Y +\sum_{j=1 \atop j\neq i}^n{N_j \over \a_i -
\a_j}, N_i].
&(2.14b)}
$$

If we now identify the flow parameters $\{t_i\}_{i=1, \dots n}$ with the
eigenvalues
$\{\a_i\}_{i=1, \dots n}$, we obtain the nonautonomous Hamiltonian systems
$$
\eqalignno{
{\di N_j\over \di \a_i}&= {[N_j,N_i]\over \a_j - \a_i}, \quad j \neq i,
 \quad i,j = 1, \dots n,  &(2.15a)
 \cr
{\di N_i\over \di \a_i} &= [Y +\sum_{j=1 \atop j\neq i}^n{N_j \over \a_i -
\a_j}, N_i],
&(2.15b)}
 $$
which are the $\a_i$ components of the system (1.1). Viewing the $N_i$'s
as functions on the fixed phase space $M$, eqs.~(2.15a,b) are induced by the
nonautonomous Hamiltonian systems generated by the pullback of the $H_i$'s
under
the parameter dependent moment map $\wt{J}_A$. Eqs.~(2.15a,b) are equivalent to
replacing the Lax equations (2.12) by the system:
$$
{\di {\cal N} \over \di \a_i} = -[(dH_i)_-,{\cal N}]
 -{\di (dH_i)_- \over \di\l\quad}, \eqno(2.16)
$$
which is just the condition of commutativity of the system of operators
$\{{\cal
D}_\l, {\cal D}_i\}_{i=1, \dots n}$, where ${\cal D}_{\l}$ is given by (1.3)
and
$$
{\cal D}_i := {\di \over \di \a_i} + (dH_i)_-(\l) ={\di \over \di \a_i}
 + {N_i \over \l - \a_i}.  \eqno(2.17)
$$
\vskip 0pt \noindent {\it Remark.}  The system (2.16) could also be viewed as a
Lax
equation defined on the dual of the centrally extended loop algebra
$\wt{gl}(r)^{\wedge}$,
in which the $\Ad^*$ action is given by gauge transformations rather than
conjugation
{\bf [RS]}. The analogue of the spectral ring $\IYA$ is the ring of monodromy
invariants,
restricted to a suitable Poisson subspace with respect to a modified
($R$--matrix) Lie
Poisson bracket structure. This viewpoint will not be developed here, but is
essential
to deriving such systems through reductions of autonomous Hamiltonian systems
of PDE's.

  The fact that the matrices $\{F_iG_i^T = L_i \in gl(k_i)\}_{i=1, \dots n}$
are constant
under the deformations generated by the eqs.~(2.15a,b), (2.16) follows from the
fact
that $$
J_{G_A}(F,G):=\diag (F_1G_1^T, \dots , F_nG_n^T) \in gl(N)  \eqno(2.18)
$$
is the moment map generating the Hamiltonian action of the stabilizer of $A$ in
$Gl(N)$:
$$
G_A := \prod_{i=1}^N Gl(k_i) =\Stab(A) \ss Gl(N),  \eqno(2.19)
$$
this action being given by
$$ \eqalignno
{G_A : M &\lra M  \cr
K: (F,G) & \lra (KF,(K^{T})^{-1}G)  &(2.20)  \cr
K = \diag (K_1, \dots ,K_n) &\in G_A , \quad  K_i\in Gl(k_i).}
$$
The orbits of $G_A$ are just the fibres of $\wt{J}_A$, so $(\wt{J}_A, J_{G_A})$
form a ``dual pair'' of moment maps {\bf [W]}. Evidently, the pullback
$\wt{J}_A^*(H)$ is $G_A$--invariant for all $H\in\IYA$, and hence $J_{G_A}$ is
constant
under the $H_i$ flows.

   So far, we have only considered the part of the system (1.1) relating to the
parameters $\{\a_i\}_{i=1, \dots n}$. What about the Hamiltonians
$\{K_a\}_{a=1, \dots r}$
that generate the $y_a$ components? As shown in {\bf [AHH1]}, besides
$\wt{J}_A$ there is, for each $Y \in gl(r)$, another moment map
$$
\eqalignno{
\wt{J}_Y : M &\lra \wt{gl}(N)_{+}^* \sim  \wt{gl}(N)_-  \cr
\wt{J}_Y : (F,G) &\lmt - F(Y -z I_r)^{-1}G^T,  & (2.21)}
$$
where $z$ denotes the loop parameter for the loop algebra $\wt{gl}(N)$, whose
elements
are defined on a circle $S^1$ in the complex $z$-plane containing the
eigenvalues
of $Y$ in its interior. The pairing identifying $ \wt{gl}(N)$ as a dense
subspace of
$\wt{gl}(N)^*$  is defined  similarly to (2.3), for elements $X_1 \in
\wt{gl}(N)^*,
\ X_2 \in  \wt{gl}(N)$. The subalgebras $\wt{gl}(N)_{\pm}$ are similarly
defined with
respect to this circle, and their dual spaces $\wt{gl}(N)_{\pm}^*$
are identified analogously with $\wt{gl}(N)_{\mp}$.

The moment map $\wt{J}_Y$ is also ``dual'' to $\wt{J}_A$, but in a different
sense than
$J_{G_A}$ --  one that is relevant for the construction of the remaining
Hamiltonians
$\{K_a\}_{a=1, \dots r}$. The space $\gA$ may be identified with the quotient
Poisson
manifold $M/G_A$, with symplectic leaves given by the level sets of the
symmetric
invariants formed from each $F_iG_i^T$, since these are the Casimir invariants
on $\gA$.
Since the Hamiltonians in $\IYA$ are all also invariant under the action of the
stabilizer
$G_Y= Stab(Y) \ss Gl(r)$, where $Gl(r) \ss \wt{Gl}(r)$ is the subgoup of
constant loops, we may also quotient by this action to obtain
$ \gA/G_Y = M /(G_Y \times G_A)$. Doing this in the opposite order, we may
define
$\gY \ss\wt{gl}(N)_{+}^* \sim \wt{gl}(N)_-$ as the Poisson subspace consisting
of
elements of the form:
 $$
\eqalignno{
{\cal M}_0(z) &= -F(Y-zI_N)^{-1}G^T=\sum_{a=1}^r {M_a\over{z-y_a}},  & (2.22a)
\cr
(M_a)_{ij}&:= F_{ia}G_{ja}, \quad i,j =1, \dots ,n, \quad a=1, \dots r
&(2.22b)}
$$
(where, if the $\{y_a\}_{a=1, \dots r}$ are distinct, the residue matrices are
all of rank
$1$), and identify $\gY$ with $M/G_Y$. Denoting by $\IAY$ the ring of
$\Ad^*$-invariant
polynomials on $\wt{gl}(N)^*$, restricted to the affine subspace $-A+ \gY$
consisting of
elements of the form
$$
{\cal M} = - A + {\cal M}_0, \quad {\cal M}_0\in \gY,  \eqno(2.23)
$$
the pullback of the ring $\IAY$ under the moment map $\wt{J}_Y$ also gives a
Poisson commuting ring whose elements are both $G_Y$ and $G_A$ invariant, and
hence
project to $ M/(G_Y \times G_A)$. In fact, the two rings $\wt{J}_A^*(\IYA)$ and
$\wt{J}_Y^*(\IAY)$ {\it coincide} (cf. {\bf [AHH1]}), because of the identity:
$$
\det(A-\l I_N) \det ( Y + G^T(A-\l I_N)^{-1}F  -z I_r) =
\det(Y - zI_r) \det (A + F(Y-zI_r)^{-1}G^T - \l I_N),  \eqno(2.24)
$$
which shows that the spectral curve of ${\cal M}(z)$, defined by
$$
\det (A + F(Y-zI_r)^{-1}G^T - \l I_N) =0  \eqno(2.25)
$$
and that of ${\cal N}(\l)$, given by eq.~(2.9), are birationally equivalent
(after reducing out the trivial factors $\det (A-\l I_N)$ and $\det (Y- z
I_r)$).

    Now, similarly to the definition of the elements $\{ H_i \in \IYA \}_{i=1,
\dots n}$,
we may define $\{ K_a \in \IAY\}_{a=1, \dots r}$ as:
$$
K_a := {1\over 4 \pi i} \oint_{z=y_a} \tr ({\cal M}(z))^2dz
= -\tr(A M_a) + \sum_{b=1 \atop b\ne a}^r {\tr(M_a M_b) \over y_a -y_b}
\eqno(2.26)
$$
To verify that these coincide with the $K_a$'s defined in eq.~(1.8b), we use
eqs.~(2.4a),
(2.8), (2.22a) and (2.23) to express $K_a$ as:
$$
\eqalignno{
K_a &= {1\over 4 \pi i} \oint_{z=y_a}dz {1\over 2\pi i}
\oint_{S^1} d\l \l \tr({\cal M}(z) (A-\l I_N)^{-1})^2 &(2.27a)\cr
&={1\over 4 \pi i}\oint_{S^1} d\l \l {1\over 2\pi i}  \oint_{z=y_a}dz
\left[\tr((Y-zI_r)^{-1}{\cal N}(\l))^2
-2\tr\left((Y-z I_r)^{-1} {\cal N}(\l)\right) \right]. &(2.27)}
$$
Evaluating residues at $\{y_a\}_{a=1,\dots r}$ in $z$ and at $\infty$ in $\l$
gives
(1.8b).  The Poisson commutativity of the $K_a$'s again follows from the AKS
theorem, and
the commutativity with the $H_i$'s follows from the equality of the two rings
$\wt{J}_A^*\IYA = \wt{J}_Y^*\IAY$. To compute the Lax form of the equations of
motion
generated by the $K_a$'s, we evaluate their differentials, viewing them as
functions of
${\cal N}(\l)$ defined by eq.~(2.27). Evaluating the $z$ integral, this gives:
$$
dK_a(\l) = \l E_a + \sum_{b=1 \atop b\ne a}^r {E_a {\cal N}(\l) E_b + E_b {\cal
N}(\l) E_a
\over y_a -y_b} \in \wt{gl}(r) \eqno(2.28)
$$
where $E_a$ denotes the elementary diagonal $r\times r$ matrix with $(aa)$
entry equal
to $1$ and zeroes elsewhere. Taking the projection to $\wt{gl}(r)_+$ gives:
$$
(dK_a)_+(\l) = \l E_a + \sum_{b=1 \atop b\ne a}^r {\sum_{i=1}^n }
{E_a N_i E_b + E_bN_i E_a \over y_a -y_b} \in \wt{gl}(r)_+, \eqno(2.29)
$$
and hence
$$
\sum_{a=1}^r (dK_a)_+dy_a = \l Y + \Theta, \eqno(2.30)
$$
where $\Theta$ is defined in eq.~(1.2). By the AKS theorem, the autonomous form
of the
equations of motion is
$$
{\di {\cal N}\over \di \tau_a} = \left[ (dK_a)_+, {\cal N}\right], \eqno(2.31)
$$
while the nonautonomous version is
$$
{\di {\cal N}\over \di y_a} = \left[ (dK_a)_+, {\cal N}\right]
+{\di (dK_a)_+ \over \di \l\quad}
 = \left[ (dK_a)_+, {\cal N}\right]+ E_a.  \eqno(2.32)
$$
Evaluating residues at $\l =\a_i$ gives the equations
$$
{\di N_i\over \di \tau_a} = \left[ (dK_a)_+(\a_i), N_i\right], \quad i=1,
\dots, n
\eqno(2.33)
 $$
for the autonomous case and
$$
{\di N_i\over \di y_a} = \left[ (dK_a)_+(\a_i), N_i\right], \quad i=1, \dots, n
\eqno(2.34)
$$
for the nonautonomous one. Eqs.~(2.34) are just the $y_a$ components of the
system (1.1).
Eq.~(2.32) is equivalent to the commutativity of the operators $\{{\cal D}_\l,
 {\cal D}_a^*\}_{a=1,\dots r}$, where
$$
{\cal D}_a^* := {\di \over \di y_a} - (dK_a)_+(\l),  \eqno(2.35)
$$
and implies that the monodromy of ${\cal D}_\l$ is invariant under the $y_a$
deformations. In fact, it follows from the AKS theorem that the complete system
of
operators $\{{\cal D}_{\l}, {\cal D}_{i},{\cal D}_a^*\}_{i=1, \dots n, a=1,
\dots r}$
commutes.

    Turning now to the dual system, it follows from the AKS theorem that the
Lax form  of
the equations of motion induced by the $K_a$'s on $\wt{gl}(N)_-$, viewed now as
functions of ${\cal M}$,  in the autonomous case is
$$
{\di{\cal M}\over \di\tau_a} = -[(dK_a)_-, {\cal M}],  \eqno(2.36)
$$
where
$$
(dK_a)_-(z) =  {M_a\over z-y_a} \in \wt{gl}(N)_-.  \eqno(2.37)
$$
(Note that the differential $dK_a$ entering in eqs.~(2.36), (2.37) and below
has a
different significance from that appearing in eqs.~(2.28)--(2.35).) Evaluating
residues at $z=y_a$ shows that (2.36) is equivalent to the system
$$
\eqalignno{
{\di M_b\over \di \tau_a}&= {[M_b,M_a]\over y_b - y_a}, \quad b \ne a,\quad
a,b = 1, \dots r,   &(2.38a) \cr
{\di M_a\over \di \tau_a} &= [-A +\sum_{b=1 \atop b\ne a}^n {M_b \over y_a -
y_b}, M_a].
 & (2.38b) }
$$
   Identifying the flow parameters $\{\tau_a\}_{a=1, \dots r}$ now with the
eigenvalues
$\{y_a\}_{a=1, \dots r}$ of $Y$ gives the nonautonomous Hamiltonian system
$$
\eqalignno{
{\di M_b\over \di y_a}&= {[M_b,M_a]\over y_b - y_a}, \quad b \ne a,\quad
a,b = 1, \dots r,   &(2.39a)
 \cr
{\di M_a\over \di y_a} &= [-A +\sum_{b=1 \atop b\ne a}^n {M_b \over y_a - y_b},
M_a]. &
(2.39b) }
$$
or, equivalently,
$$
{\di {\cal M} \over \di y_a} = -[(dK_a)_-,{\cal M}] -{\di (dK_a)_- \over \di z
\quad}.
\eqno(2.40)
$$
Equations (2.39a,b), (2.40) are equivalent to the commutativity of the system
of
operators $\{ {\cal D}_z, {\cal D}_a\}_{a=1, \dots r}$ defined by:
$$ \eqalignno{
 {\cal D}_z & := {\di \over \di z} - {\cal M}(z),  & (2.41a) \cr
{\cal D}_a &:= {\di \over \di y_a} + (dK_a)_-(z) =
{\di \over \di y_a} + {M_a \over z - y_a}, \quad a=1, \dots ,r.
 &(2.41b)}
$$
Thus the monodromy of the ``dual'' operator ${\cal D}_z$, is also preserved
under the $y_a$
deformations.

     Finally, by similar computations to the above, with $(A,\l, {\cal N})$
and  $(Y, z, {\cal M})$ interchanged,  we can express the $H_i$'s as functions
of
${\cal M}$:
$$
H_i ={1\over 4 \pi i}\oint_{s^1} dz z {1\over 2\pi i}  \oint_{\l=\a_i}d\l
\left[\tr((A-\l I_N)^{-1}{\cal M}(z))^2
-2\tr\left((A-\l I_N)^{-1} {\cal M}(z)\right) \right]. \eqno(2.42)
$$
Evaluating residues at $\{\a_j\}_{j=1,\dots n}$ in $\l$ and at $\infty$ in $z$
gives
$$
 H_i = -\sum_{a=1}^r(M_a)_{ii} +
\sum_{j=1 \atop j \ne i}^n {( \sum_{a=1}^r(M_a)_{ij}) ( \sum_{b=1}^r(M_b)_{ji})
\over \a_i
- \a_j}, \quad  1=1, \dots, n. \eqno(2.43)
$$
To compute the Lax equations for ${\cal M}(z)$ generated by the
$H_i$'s, we evaluate their differentials when viewed as functions on
$\wt{gl}(N)_-$ defined by eq.~(2.42). Evaluating the $\l$ integral, this gives:
$$
dH_i(z) = -z E_i + \sum_{j=1 \atop j\ne i}^r {E_i {\cal M}(z) E_j + E_j {\cal
M}(z) E_i
\over \a_i -\a_j} \in \wt{gl}(N), \eqno(2.44)
$$
where $E_i$ now denotes the elementary diagonal $N \times N$ matrix with $(ii)$
entry
equal to $1$ and zeroes elsewhere. Taking the projection to $\wt{gl}(N)_+$
gives:
$$
(dH_i)_+(z) = -z E_i + \sum_{j=1 \atop j\ne i}^n {\sum_{a=1}^r }
{E_i M_a E_j + E_jM_a E_i \over \a_i - \a_j} \in \wt{gl}(N)_+, \eqno(2.45)
$$
and hence
$$
\sum_{i=1}^n (dH_i)_+(z)d\a_i = -z dA + \Phi, \eqno(2.46)
$$
where
$$
\Phi_{ij}=(1-\d_{ij})(\sum_{a=1}^r(M_a)_{ij} d\log(\a_i - \a_j).  \eqno(2.47)
$$
 By the AKS theorem, the autonomous form of the equations of motion is
$$
{\di {\cal M}\over \di t_i} = \left[ (dH_i)_+, {\cal M}\right] \eqno(2.48)
$$
while the nonautonomous system is
$$
{\di {\cal M}\over \di \a_i} = \left[ (dH_i)_+, {\cal M}\right]
+{\di (dH_i)_+ \over \di z\quad}
= \left[ (dH_i)_+, {\cal M}\right] - E_i.
\eqno(2.49)
$$
Evaluating residues at $\l =\a_i$ gives the equations
$$
{\di M_a\over \di t_i} = \left[ (dH_i)_+(y_a), M_a\right] \eqno(2.50)
$$
for the autonomous case and
$$
{\di M_a\over \di \a_i} = \left[ (dH_i)_+(y_a), M_a\right]  \eqno(2.51)
$$
for the nonautonomous one. Eq.~(2.49) is equivalent to the commutativity of the
operators
$\{{\cal D}_z, {\cal D}_i^*\}_{i=1,\dots n}$, where
$$
{\cal D}_i^* := {\di \over \di \a_i} - (dH_i)_+(z),  \eqno(2.52)
$$
and implies that the monodromy of ${\cal D}_z$ is invariant under the $\a_i$
deformations. Again, it follows from the AKS theorem that the complete system
of
operators $\{{\cal D}_{z}, {\cal D}_{a},{\cal D}_i^*\}_{a=1, \dots r, i=1,
\dots n}$
commutes.

   Thus, at the level of the reduced spaces $\gA/G_Y \sim \gY/G_A$, we have two
equivalent ``dual'' isomonodromy representations of the Hamiltonian systems
generated by
the spectral invariants $\{H_i, K_a\}_{i=1, \dots ,n, a=1,\dots ,r}$ -- systems
(2.15a,b),
(2.24) in the ${\cal N}(\l) \in \gA$ representation and (2.39a,b), (2.51) in
the
${\cal M}(z) \in \gY$ representation.
 %
\bigskip \noindent {\bf 3.  Reductions}\hfill

  The general scheme of {\bf [AHP, AHH1]} may combined with
continuous or discrete Hamiltonian symmetry reductions to deduce systems
corresponding
to subalgebras of $\wt{gl}(r)$ and $\wt{gl}(N)$ or, more generally, to
invariant
submanifolds. In particular, the Marsden-Weinstein reduction method may be
applied to the
symmetry groups $G_A$ and $G_Y$, or to other invariants of the system.

	The discrete reduction method (see {\bf [AHP], [HHM]} for further details) may
be
summarized as follows. Suppose $\wt{\s}_{r*} :\wt{gl}(r)_+ \ra \wt{gl}(r)_+$ is
a
Lie algebra homomorphism that is semisimple, of finite order and induced by the
group
homomorphism $\wt{\s}_r:\wt{Gl}(r)_+ \ra \wt{Gl}(r)_+$. Let
$\wt{\s}_{r}^* :\wt{gl}(r)_-\ra \wt{gl}(r)_-$ denote the dual map, which is a
Poisson homomorphism, and let $h_{\s} \ss \wt{gl}(r)_+$, $h_{\s}^* \ss
\wt{gl}(r)_-$
denote the subspaces consisting of the fixed point sets under $\wt{\s}_{r*}$
and
$\wt{\s}_{r}^*$, respectively. (These are naturally dual to each other, since
$h_{\s}^* $
may be identified with the annihilator of the complement of $h_{\s}$ under the
decomposition of $\wt{gl}(r)_+$ into eigenspaces of $\wt{\s}_{r*}$). Then
$h_{\s} \ss
\wt{gl}(r)_+$ is a subalgebra, and  its dual space $h_{\s}^*$ has the
corresponding Lie
Poisson structure. Suppose there also exists a finite order symplectomorphism
$\s_M:M\ra
M$ such that the moment map $\wt{J}_A$ satisfies the intertwining property:
$$
\wt{J}_A \circ \s_M = \wt{\s}_r^* \circ \wt{J}_A.  \eqno(3.1)
$$
The fixed point set $M_\s \ss M$ is, generally, a symplectic submanifold,
invariant under
the flows generated by $\s_M$-invariant Hamiltonians. The restriction
$\wt{J}_A\vert_{M_\s} := \wt{J}_{A\s}$ takes its values in $h_{\s}^*$, defining
a Poisson
map: $$
\wt{J}_{A\s} :M_{\s} \lra h_{\s}^*, \eqno(3.2)
$$
which is the equivarient moment map generating the action of the subgroup $H_\s
\ss\wt{Gl}(r)_+$ consisting of the fixed points under  $\wt{\s}_r:\wt{Gl}(r)_+
\ra
\wt{Gl}(r)_+$. Such reductions, when applied to the spectral invariants on $Y
+\gA$
generate systems satisfying the criteria of the AKS theorem, provided the
matrices $Y$ and
$A$ are appropriately chosen to be compatible with the reduction. The same
procedure may
be applied to the dual systems on $-A+\gY$ if a similar homomorphism
$\wt{\s}_{N*}:
\wt{gl}(N)_+ \ra \wt{gl}(N)_+$ exists, satisfying the intertwining property:
$$
\wt{J}_Y \circ \s_M = \wt{\s}_N^* \circ \wt{J}_Y.  \eqno(3.3)
$$
The corresponding moment map,
$$
\wt{J}_{Y\s} :M_{\s} \lra k_{\s}^*, \eqno(3.4)
$$
 obtained by restriction $\wt{J}_{Y\s} :=\wt{J}_Y\vert_{M_\s}$ takes its values
in
the Poisson subspace $k_{\s}^* \ss \wt{gl}(N)_-$ consisting of the fixed point
set under
the dual  map $\wt{\s}_N^*:\wt{gl}(N)_- \ra \wt{gl}(N)_-$, and $k_{\s} \ss
\wt{gl}(N)_+$
is the corresponding  subalgebra consisting of fixed points under
$\wt{\s}_{N*}$.

  The following examples illustrate both the discrete and continuous reduction
procedures.
\medskip \noindent {\it 3a.  Symplectic Reduction (discrete)}\hfill

Let $r=2s$ and define $\wt{\s}_r^*, \wt{\s}_N^*$ and $\s_M$ by:
$$
\eqalignno{
\wt{\s}_r^*: X(\l) &\lra J X^T(\l)J,   &(3.5a)  \cr
\wt{\s}_N^*: \xi(z) &\lra  \xi^T(-z),   &(3.5b)  \cr
\s_M: (F, G) &\lra J (GJ, -FJ),   &(3.5c)  \cr
X \in \wt{gl}(r)_-, &\quad \xi \in \wt{gl}(N)_-,\quad (F,G) \in M}
$$
where
$$
J = \pmatrix {0 & I_s \cr
  -I_s & 0 } \in M^{2s\times 2s}.
  \eqno(3.6)
$$
Then $M_\s \ss M$ consists of pairs $(F,G)$ of the form:
$$
F= {1\over \sqrt{2}}(Q, P), \quad  G ={1\over \sqrt{2}}(P, -Q)  \eqno(3.7)
$$
where $Q,P \in M^{N\times s}$ are $N\times s$ matrices. The blocks
$\{(F_i, G_i)\}_{i=1, \dots n}$ corresponding to the eigenvalues
$\{\a_i\}_{i=1, \dots n}$
are similarly of the form
$$
F_i= {1\over \sqrt{2}}(Q_i, P_i), \quad  G_i ={1\over \sqrt{2}}(P_i, -Q_i),
\eqno(3.8)
$$
where $Q_i,P_i \in M^{k_i\times s}$.
The reduced symplectic form on $M_\s$ is
$$
\omega = \tr(dQ \wedge dP^T).  \eqno(3.9)
$$
The subalgebra $h_\s \ss \wt{gl}(r)_+$ is just the positive part
$\wt{sp(2s)}_+$ of the
symplectic loop algebra $\wt{sp(2s)}$, and the dual space $h_\s^*$ is similarly
identified
with  $\wt{sp(2s)}_-$. The subalgbra $k_\s \ss \wt{gl}(N)_+$ is the positive
part
$\wt{gl}^{(2)}(N)_+$ of the ``twisted'' loop algebra $\wt{gl}^{(2)}(N)$, with
dual
space $k_s^* \sim \wt{gl}^{(2)}(N)_-$. The image of the moment map
$\wt{J}_{A\s}$ has the
form
$$
{\cal N}_0(\l) = \wt{J}_{A\s}(Q,P) =  \sum_{i=1}^n{\pmatrix{
 \scst -P_i^TQ_i &\scst -P_i^TP_i  \cr
\scst Q_i^TQ_i & \scst Q_i^TP_i}\over \l - \a_i}.  \eqno(3.10)
$$
In order that the pullback of the elements of the ring $\IYA$ under $\wt{J}_A$
be
$\s_M$-invariant, and that the intertwining property (3.3) for the moment map
$\wt{J}_Y$ be
satisfied, the matrix $Y$ must be in $sp(2s)$. For diagonal $Y$, this means
$$
\eqalignno{
Y &= \pmatrix{ y & 0 \cr
                0 & - y}, & (3.11) \cr
y & =\diag(y_1, \dots \ y_s) \in gl(s).}
$$
The image of  $\wt{J}_{Y\s}$ then has the block form
$$
{\cal M}_0(z) = \wt{J}_{Y\s}(Q,P) = {1\over 2} \sum_{a=1}^r \left( {q_a p_a^T
\over z-y_a}
- {p_a q_a^T \over z+y_a}\right), \eqno(3.12)
$$
where $\{q_a\}_{a=1, \dots r}$ and  $\{p_a\}_{a=1, \dots r}$ denote the $a$th
columns of
$Q$ and $P$, respectively.

   The Hamiltonians $\{H_i\}_{i=1, \dots n}$ reduce in this case to:
$$
H_i = -\tr (y P_i^TQ_i) +{1\over 4} \sum_{j=1 \atop j\ne i}^n{\tr\left(
(Q_iP_j^T
-P_iQ_j^T)(P_jQ_i^T -Q_jP_i^T)\right)\over \a_j - \a_i}, \eqno(3.13)
$$
and generate the equations of motion:
$$
\eqalignno{
{\di Q_i \over \di \a_i}&= {P_iQ_j^TQ_j -Q_iP_j^TQ_j\over 2(\a_i -\a_j)}, \quad
j \ne i
&(3.14a) \cr
{\di P_i \over \di \a_i}&= {P_iQ_j^TP_j -Q_iP_j^TP_j\over 2(\a_i -\a_j)}, \quad
j \ne i
&(3.14b) \cr
{\di Q_i \over\di \a_i}& =- Q_i y + \sum_{j=1 \atop j\ne i}^r {Q_iP_j^TQ_j -
P_i Q_j^TQ_j
                           \over 2(\a_i -\a_j)} &  (3.14c)\cr
{\di P_i \over\di \a_i}& = P_i y + \sum_{j=1 \atop j\ne i}^r{Q_iP_j^TP_j
-P_iQ_j^TP_j
\over 2(\a_i -\a_j)}. &  (3.14d)  }
$$
  The Hamiltonians $\{K_a= -K_{a+s}\}_{a=1, \dots s}$ reduce to:
$$
K_a ={1\over 2} p_a^T Aq_a +{1\over 4} \sum_{b=1 \atop b\ne a}^r {\tr
(q_ap_a^Tq_bp_b^T)
\over y_a -y_b}-{1\over 4}\sum_{b=1}^r {\tr (q_a p_a^Tp_bq_b^T) \over y_a +y_b}
\eqno(3.15)
$$
and generate the equations
$$
\eqalignno{
{\di q_a \over \di y_b}&= {1\over 4}{q_bp_b^Tq_a \over y_b -y_a} +
                       {1\over 4}{p_bq_a^Tq_b \over y_b +y_a}, \quad b\ne a &
(3.16a) \cr
{\di p_a \over \di y_b}&= -{1\over 4}{p_bp_a^Tp_b \over y_b -y_a} -
                       {1\over 4}{q_bp_b^Tp_a \over y_b +y_a}, \quad b\ne a &
(3.16b) \cr
{\di q_a \over \di y_a} &= {1\over 2}Aq_a +
       {1\over 4} \sum_{b=1 \atop b\ne a}^r{q_bp_b^Tq_a \over y_a -y_b} +
       {1\over 4} \sum_{b=1 \atop b\ne a}^r{p_bq_b^Tq_a \over y_b +y_a}  &
(3.16c) \cr
{\di p_a \over \di y_a} &= -{1\over 2}Ap_a -
       {1\over 4} \sum_{b=1 \atop b\ne a}^r{p_bp_a^Tq_b \over y_a -y_b} +
       {1\over 4} \sum_{b=1 \atop b\ne a}^r{q_bp_a^Tp_b \over y_b +y_a}.  &
(3.16d) }
$$

The particular case $s=1,\{ k_i=1\}_{i=1, \dots n}$ of (3.13), (3.14a-d)
reduces, up to a
simple change of basis, to the system of Theorem 7.5,  {\bf[JMMS]}. The
corresponding
$\tau$-function gives the $n$-particle correlation function for an impenetrable
Bose gas
or, equivalently, the level spacing distribution function for a set of random
matrices
having no eigenvalues in the intervals $\{[\a_{2i-1},\a_{2i}]\}_{i=1,\dots
m},\ n=2m$,
in the scaling limit {\bf [TW1]} .
\medskip \noindent
{\it Hamiltonian Structure of Painlev\'e Equations}

The following two examples show how the Painlev\'e transcendents
$P_V$ and $P_{VI}$  may be derived from the generic systems (2.15a,b), (2.34)
or
(2.39a,b), (2.51) through Hamiltonian reduction under continuous symmetry
groups. Our
derivation will be guided by the formulation of Painlev\'e transcendents as
monodromy
preserving deformation equations given in {\bf [JM]}, but the emphasis here
will be on the
loop algebra content, the Hamiltonian reductions and the associated ``dual''
systems. For
previous work on the Hamiltonian structure of the Painlev\'e transcendents, see
{\bf[Ok]}
and references therein.

\noindent {\it 3b.  Painlev\'e V}\hfill

Choose $N=2$, $r=2$, and
$$
Y= \pmatrix{
t & 0\cr
0 & -t}, \quad
A= \pmatrix{
0 & 0\cr
0 & 1}. \eqno(3.17)
$$
Then $F$ and $G$ are $ 2 \times 2$ matrices
$$
F =\pmatrix{
 F_1 \cr
F_2 }, \quad
G = \pmatrix{
G_1 \cr
G_2 }, \quad  \eqno(3.18)
$$
with row vectors  $\{F_i = (F_{i1}\ F_{i2}),\  G_i = (G_{i1}\ G_{i2})\}_{ i=1,
2}$.
The stabilizer $G_A \ss Gl(2)$ of $A$
is the diagonal subgroup generated by the moment map
$$
J_{G_A}(F,G) = (F_1G_1^T,  F_2G_2^T) :=(\mu_1, \mu_2). \eqno(3.19)
$$
Fixing a level set, we parametrize the quotient under this abelian
Hamiltonian group action by choosing the symplectic section $M_A \ss M$ defined
(on a
suitable connected component) by
$$
F = {1\over \sqrt{2}}
\pmatrix{
x_1 & y_1 -{\mu_1 \over x_1}  \cr
x_2 & y_2 -{\mu_2 \over x_2}}, \quad
G = {1\over \sqrt{2}}
\pmatrix{
y_1+ {\mu_1 \over x_1} & -x_1 \cr
y_2+ {\mu_2 \over x_2} & -x_2 }.\eqno(3.20)
$$
The reduced manifold
$M_{\red} = J_{G_A}^{-1}(\mu_1,\mu_2) / G_A$ is
identified with ${\bf R}^2 \times {\bf R}^2$ minus the
coordinate axes $\{x_1=0, x_2=0\}$, quotiented by the group of
reflections in these axes. The reduced symplectic form is
$$
\o_{\red} = \sum_{i=1}^2 dx_i\wedge dy_i. \eqno(3.21)
$$
 The image of the reduced moment map
$\wt{J}_A:M_{\red} \ra \wt{gl}(2)_-$ translated by $Y$ is
$$
\eqalign{
{\cal N}(\l) &= Y+\wt{J}_{A}(F,G) \cr
           &=  \pmatrix{t & 0 \cr
               0  & -t} +
{ \pmatrix{\scst -x_1 y_1 - \mu_1 &\scst -y_1^2 + {\mu_1^2\over x_1^2} \cr
     \scst   x_1^2  &\scst  x_1y_1 - \mu_1} \over 2\l} +
{ \pmatrix{\scst -x_2 y_2 - \mu_2 &\scst -y_2^2 + {\mu_2^2\over x_2^2} \cr
      \scst  x_2^2  &\scst  x_2y_2 - \mu_2} \over 2(\l - 1)}. } \eqno(3.22)
$$
The stabilizer $G_Y \ss Gl(2)$ of $Y$ is the diagonal subgroup, acting by
conjugation on ${\cal M}(\l)$, which corresponds to scaling transformations
$$
(x_1,x_2,y_1,y_2) \lra (e^{\t} x_1, e^{\t} x_2, e^{-\t} y_1, e^{-\t} y_2),
\eqno(3.23)
$$
and is generated by
$$
 a:= {1\over 2}(x_1y_1+x_2 y_2). \eqno(3.24)
$$
(The trace part of $gl(2)$ acts trivially, since it coincides with the
Casimir $\mu_1+\mu_2$.)

The $G_Y$--invariant Hamiltonians $H_1, H_2 \in \IYA$ may be expressed
$$
\eqalignno{ H_1 = &{1\over 4 \pi i} \oint_{\l=0} \tr({\cal N}(\l))^2 d\l
= \tr(Y N_1)  - \tr(N_1 N_2)  = -tH - {\mu_1 \mu_2 \over 2} -a^2 -2at &(3.25a)
\cr
 H_2 = &{1\over 4 \pi i} \oint_{\l=1} \tr({\cal N}(\l))^2 d\l
= \tr(Y N_2) + \tr(N_1 N_2)  = tH + {\mu_1 \mu_2 \over 2} +a^2. &(3.25b) }
$$
where
$$
H = -{1\over 4t}(x_1^2 + x_2^2)( y_1^2 + y_2^2) + {1\over 4t}\left(
\mu_1^2{ x_2^2 \over x_1^2} +\mu_2^2{ x_1^2 \over x_2^2}\right) -x_2 y_2.
\eqno(3.26)
$$

The dual system is determined by the moment map $\wt{J}_Y:M \ra \wt{gl}(2)$
defined by eq.~(2.21) which, when restricted to the symplectic section $M_A$
defined by
eq.~(3.20), gives
$$
\eqalignno{
{\cal M}(z)=& - A - F(Y-zI_2)^{-1}G^T  \cr
=&\pmatrix{ 0 & 0\cr 0 & -1} +
{\pmatrix{\scst x_1 y_1 +  \mu_1 &\scst x_1 y_2 + \mu_2{ x_1 \over x_2} \cr
\scst x_2 y_1 + \mu_1 {x_2 \over x_1} &\scst x_2y_2 +  \mu_2}\over 2(z-t)} +
{\pmatrix{\scst -x_1 y_1 +  \mu_1 &\scst - x_2 y_1 + \mu_1{ x_2 \over x_1} \cr
\scst - x_1 y_2 + \mu_2 {x_1 \over x_2} &\scst -x_2y_2 +  \mu_2}\over 2(z+t)}.
&(3.27)}
$$
Here, the quantities $a$ and $\mu_1+\mu_2$ are interpreted as Casimir
invariants,  whereas $\mu_1$ and $\mu_2$ individually are conserved
quantities because they belong to the spectral ring $\IAY$. In terms of the
dual system,
we may express the Hamiltonians $K_1, K_2 \in \IAY$ as:
$$
\eqalignno {K_1 &= {1\over 4 \pi i} \oint_{\l=t} \tr({\cal M}(z))^2 dz
= -\tr(A M_1) +{1\over 2t} \tr(M_1 M_2) \cr
&=\quad  {H\over 2} + {\mu_1^2 +\mu_2^2 \over 8t}
  -{\mu_2\over 2} &(3.28a) \cr
K_2 &= -{1\over 4 \pi i} \oint_{\l=t} \tr({\cal M}(z))^2 dz
= -\tr(A M_2) -{1\over 2t} \tr(M_1 M_2) \cr
&= -{H\over 2} - {\mu_1^2 +\mu_2^2 \over 8t}
 -{\mu_2 \over 2}. &(3.28b)}
 $$
The relations
 $$
\eqalignno{
H_1 &= -H_2 -2at & (3.29a)\cr
& = -2tK_1 + {(\mu_1 - \mu_2)^2 \over 4} - a^2 - 2 a t - t \mu_2 &(3.29b) \cr
& = \phantom{-}2tK_2 + {(\mu_1 - \mu_2)^2 \over 4} - a^2 - 2 a t + t \mu_2 &
(3.29c)}
$$
can also be derived from the identity (following from  eq.~(2.24)):
$$
{ z^2 - z\tr{\cal N}(\l) + {1\over 2} \left((\tr {\cal N}(\l))^2-
\tr{\cal N}^2(\l)\right)\over z^2 -t^2}
= { \l^2 + \l\tr{\cal M}(z) + {1\over 2} \left((\tr {\cal M}(z))^2-
\tr{\cal M}^2(z)\right)\over\l(\l-1)}. \eqno(3.30)
$$
Integrating both sides around contours in the $z$-plane containing either the
pole at
$z=t$ or the one at $z=-t$ and contours in the $\l$-plane containing either the
pole at
$\l=0$ or the one at $1$, we obtain eqs.~(3.29a-c).

Viewing $K_1$ and $K_2$  as  functions of ${\cal N}$ we have, from eq.~(2.29)
 $$
\eqalignno{
(dK_1)_+& = \pmatrix{\l & 0 \cr
               0 & 0} +
         {1\over 4t} \pmatrix{\scst 0 &
      \scst - y_1^2 - y_2^2 + {\mu_1^2\over x_1} +  {\mu_2^2\over x_2}\cr
            \scst x_1^2 + x_2^2 & \scst 0}   &(3.31a) \cr
(dK_2)_+& = \pmatrix{0 & 0 \cr
               0 &\l} -
         {1\over 4t} \pmatrix{\scst 0 &
      \scst - y_1^2 - y_2^2 +  {\mu_1^2\over x_1} +  {\mu_2^2\over x_2}\cr
            \scst x_1^2 + x_2^2 & \scst 0}&(3.31b) \cr
(dK_1)_+ - (dK_2)_+& = \pmatrix{\l & 0 \cr
               0 &-\l} +
         {1\over 2t} \pmatrix{\scst 0 &
    \scst - y_1^2 - y_2^2 +  {\mu_1^2\over x_1} +  {\mu_2^2\over x_2}\cr
            \scst x_1^2 + x_2^2 & \scst 0},  &(3.31c)  }
$$
and hence the monodromy preserving deformation equation for ${\cal N}(\l)$
generated by
$K_1 -K_2$ is given by the commutativity of the operators ${\cal D}_\l ={\di
\over \di \l}
-{\cal N}(\l)$ and ${\cal D}_t^*$, where  ${\cal N}(\l)$ is given by
eq.~(3.22), and
$$
{\cal D}_t^*= {\di \over \di t}- \pmatrix{\l & 0 \cr
                                   0 & -\l}  -
            {1\over 2t} \pmatrix{\scst 0 &
             \scst - y_1^2 - y_2^2 + \mu_1 {y_1\over x_1} + \mu_2 {y_2\over
x_2}\cr
            \scst x_1^2 + x_2^2 & \scst 0} . \eqno(3.32)
$$

Viewing $K_1$ and $K_2$ instead as  functions of ${\cal M}$, from (2.37) we
have
$$
\eqalignno{
(dK_1)_- &= {\pmatrix{\scst x_1 y_1 +  \mu_1 &\scst x_1 y_2 + \mu_2{ x_1 \over
x_2} \cr
\scst x_2 y_1 + \mu_1 {x_2 \over x_1} &\scst x_2y_2 +  \mu_2}\over 2(z-t)}
&(3.33a) \cr
(dK_2)_- & = {\pmatrix{\scst- x_1 y_1 +  \mu_1 &\scst - x_2 y_1 + \mu_1{ x_2
\over x_1} \cr
\scst- x_1 y_2 + \mu_2 {x_1 \over x_2} &\scst -x_2y_2 +  \mu_2}\over 2(z+t)}.
&(3.33b)}
$$
To obtain  the corresponding dual monodromy preserving deformation equation for
${\cal M}(z)$, we cannot simply restrict eqs.~(2.39a,b), (2.40) to the
submanifold $M_A \ss
M$. The image of $M_A$ under $\wt{J}_Y$ is $3$--dimensional, since points
related by the
scaling transformation (3.23) have the same image. Unlike
$\wt{J}_A\vert_{M_A}$, the
restriction  $\wt{J}_Y\vert_{M_A}$ is not a Poisson map, since $G_A$ does not
leave $\wt{J}_Y$ invariant, but acts by conjugation on the image
$\wt{J}_Y(F,G)$.
Therefore $\wt{J}_Y$ does not project to define a map on the quotient space
$M/G_A$.
However, the reduced systems generated by the $G_A$ and $G_Y$--invariant
Hamiltonians $H_1,
H_2, K_1$ and $K_2$ {\it are} determined by the projection of their Lax
equations to the
quotient manifold $M/G_A$. The projected (nonautonomous) Hamiltonian vector
field determined
by $K_1 -K_2$ has a unique lift that preserves the section $M_A$, obtained by
adding a
``vertical'' term   $$
dK_v :=
\pmatrix{ \scst {1\over 2t}({\mu_1 x_2^2 \over x_1^2} - \mu_2) & \scst 0 \cr
\scst 0 &\scst {1\over 2t}({\mu_2 x_1^2 \over x_2^2} - \mu_1)}  \eqno(3.34)
$$
to the factor $(dK_1)_- - (dK_2)_-$ entering in the Lax equation for ${\cal
M}(z)$.
 Eq.~(3.34) is obtained by noting that, apart from the conditions that the
diagonal terms
$\mu_1, \mu_2$ in $M_1 + M_2$ be conserved, and the Casimirs det$M_1$, det$M_2$
vanish,
all of which are automatically satisfied by the Lax system, the only remaining
condition
defining the image $\wt{J}_Y(M_A)$ is:
$$
\det\left({\cal M}(z) - {\cal M}^T(-z)\right) =0, \eqno(3.35a)
$$
or, equivalently,
$$
\det\left(M_1 + M_2^T\right) =0. \eqno(3.35b)
$$
Up to multiples of the identity matrix, the unique element  in the diagonal
subalgebra
(corresponding to $G_A$) which, when added to $(dK_1)_- - (dK_2)_-$, preserves
this
condition is $dK_v$.  Thus, the correct dual deformation operator ${\cal D}_t$,
whose
commutativity with ${\cal D}_z ={\di \over \di z} -{\cal M}(z)$, gives the
Hamiltonian
system generated by $K_1 - K_2$, is:
$$
\eqalign{
 {\cal D}_t =& {\di \over \di t} +
{\pmatrix{\scst x_1 y_1 +  \mu_1 &\scst x_1 y_2 + \mu_2{ x_1 \over x_2} \cr
\scst x_2 y_1 + \mu_1 {x_2 \over x_1} &\scst x_2y_2 +  \mu_2}\over 2(z-t)}
-{\pmatrix{\scst -x_1 y_1 +  \mu_1 &\scst - x_2 y_1 + \mu_1{ x_2 \over x_1} \cr
\scst -x_1 y_2 + \mu_2 {x_1 \over x_2} &\scst -x_2y_2 +  \mu_2}\over 2(z+t)}
\cr
& \quad \ + \pmatrix{ \scst {1\over 2t}({\mu_1 x_2^2 \over x_1^2} - \mu_2) &
\scst 0 \cr
\scst 0 &\scst {1\over 2t}({\mu_2 x_1^2 \over x_2^2} - \mu_1)}.} \eqno(3.36)
$$

    To reduce the system under the $G_Y$ action generated by $a$, we first
introduce
the ``spectral Darboux coordinates'' $(u,w)$ (cf.~{\bf[AHH2]})  defined  by
$$
2{\cal N}(\l)_{21} =  \left({x_1^2 \over \l }+ {x_2^2 \over \l -1}\right)
 := {w(\l -u) \over \l(\l-1)}, \eqno(3.37)
$$
where
$$
\eqalignno{
w & = x_1^2 +  x_2^2  & (3.38a) \cr
u &= { x_1^2\over x_1^2 +  x_2^2}.   & (3.38b)}
$$
In terms of these, the symplectic form (3.21) becomes
$$
\o_{\red} = d\log w\wedge da   + du \wedge dv, \eqno(3.39)
$$
where
$$
v = {1\over 2} \left({x_1y_1 \over u} + {x_2y_2 \over u-1}\right)  \eqno(3.40)
$$
is the momentum conjugate to $u$ and $a$ is the invariant defined in
eq.~(3.24).
The operators ${\cal D}_\l, {\cal D}_z,{\cal D}_t^*$ and ${\cal D}_t$ may be
expressed in
terms of these coordinates by substituting
$$
\eqalign{
x_1^2 &=u w, \quad x_2^2 = (1-u)w, \quad x_1y_1 =2u(a - u v + v),
 \quad x_2y_2 = 2(1-u)(a-u v) \cr
y_1^2 &= 4{u\over w}(a-u v + v)^2, \quad  y_2^2 = 4{(1-u)\over w} (a-u v)^2 }
\eqno(3.41)
$$
in eqs.~(3.22), (3.27), (3.32) and  (3.36).

Choosing a level set $a=a_0$, the symplectic form (3.21) reduces  to
 $$
\o_{\red}\vert_{a=a_0} = du \wedge dv, \eqno(3.42)
$$
so  $(u,v)$ provide canonical coordinates on the the reduced space obtained by
quotienting
by the $G_Y$-flow. From eqs.~(3.22), (3.25a,b) it follows that we may write
$$
{1\over 2} \tr {\cal N}^2(\l) = t^2 + {H_1\over \l} + {H_2 \over \l-1}
 +  {\mu_1^2 \over 2\l^2} + {\mu_2^2 \over 2(\l-1)^2}. \eqno(3.43)
$$
{}From (3.37) and (3.40) it follows that
$$
{1\over 2} \tr {\cal N}^2(u) = (v-t)^2 +{1\over 4}\left({\mu_1 \over u} +
{\mu_2\over
u-1}\right)^2.
 \eqno(3.44)
$$
Evaluating the integral
$$
\eqalign{
{1\over 4\pi i} \oint_{\l=u}{\l(\l-1)\tr{\cal N}^2(\l)\over \l-u}d\l
 &={1\over 2}u(u-1) \tr {\cal N}^2(u) \cr
&= t^2u(u-1) + u(H_1 +H_2) -H_1 + {\mu_1^2 (u-1)\over 2u} + {\mu_2^2 u\over
2(u-1)}}\eqno(3.45)
$$
and using eqs.~(3.29a-c), (3.44), (3.45) gives
$$
K_1 -K_2 = {u(u-1)\over t} (v^2 -2 v t) + 2 a u +
{\mu_1^2 \over 4 u t} -{\mu_2^2 \over 4(u-1)t} -{a^2\over t} - 2a. \eqno(3.46)
$$
(Note that the simple canonical change of coordinates $v \ra v +{\mu_1\over 2u}
+
{\mu_2 \over 2(u-1)}$ transforms this to polynomial form; cf.~{\bf [Ok]}).
 The reduced equations of motion generated by the Hamiltonian $K_1 - K_2$ are
therefore
$$
\eqalignno{
{du\over dt}& = {2u(u-1)\over t}(v-1)  & (3.47a)\cr
{dv\over dt}& = -{2u-1\over t}(v^2-2vt) +
{\mu_1^2 \over 4u^2 t} - {\mu_2^2 \over 4(u-1)^2 t} - 2a. & (3.47b)}
$$
Eliminating $v$ by taking second derivatives gives:
$$
\eqalign{
{d^2u\over dt^2} = &\left({1\over 2u} + {1\over 2(u-1)}\right) \left({du\over
dt}\right)^2
-{1\over t}{du\over dt}  - \a {u\over t^2 (u-1)}-\b {u-1\over t^2 u} \cr
&\quad -\g {u(u-1)\over t} -\d u(u-1)(2u-1)}, \eqno(3.48)
$$
where
$$
\a= {\mu_2^2\over 2}, \quad \b = -{\mu_1^2\over 2}, \quad
\g= 4a +2, \quad \d= 2. \eqno(3.49)
$$
which is one of the equivalent forms of $P_{V}$. The more usual form is
obtained by
transforming to the new variable
$$
 w={u\over u-1}.  \eqno(3.50)
$$

\medskip \noindent {\it 3c.  Painlev\'e VI}\hfill

 We take $N=3,\ r=2,\  A= \diag(0, 1, t),\ Y=0$, so $F,G$ are $3 \times 2$
matrices
$$
F =\pmatrix{ F_1 \cr F_2 \cr F_3}, \quad G= \pmatrix{ G_1 \cr G_2 \cr G_3},
\eqno(3.51)
$$
with rows $\{F_i = (F_{i1} \ F_{i2}),\  G_i = (G_{i1} \ G_{i2})\}_{ i=1, \dots
3}$.
If  $t \ne 0,1 $ the eigenvalues of $A$ are distinct and the stabilizer $G_A
\ss Gl(3)$
is the diagonal subgroup generated by the moment map
$$
J_{G_A}=(\mu_1, \mu_2, \mu_3)=
(F_1G_1^T,  F_2G_2^T ,  F_3G_3^T). \eqno(3.52)
$$
Fixing a level set, we parametrize the quotient under this abelian
Hamiltonian group action by choosing the symplectic section $M_A \ss M$ defined
(on a
suitable connected component) by
$$
F = {1\over \sqrt{2}}
\pmatrix{
x_1 & y_1 - {\mu_1 \over x_1} \cr
x_2 & y_2 - {\mu_2 \over x_2}\cr
x_3 & - y_3 + {\mu_3 \over x_3} }, \quad
G = {1\over \sqrt{2}}
\pmatrix{
 y_1+ {\mu_1 \over x_1} & -x_1 \cr
 y_2+ {\mu_2 \over x_2} & -x_2 \cr
 y_3+ {\mu_3 \over x_3} & x_3}.\eqno(3.53)
$$
(The choice of signs is made such that subsequent reductions be at nonsingular
points in
the real case). The reduced symplectic form is
$$
\o_{\red} = \sum_{i=1}^3 dx_i\wedge dy_i,  \eqno(3.54)
$$
and the reduced manifold $M_{\red} = J_{G_A}^{-1}(\mu_1,\mu_2,\mu_3) / G_A$ is
identified with ${\bf R}^3 \times {\bf R}^3$ minus the
coordinate planes $\{x_1=0, x_2=0, x_3=0\}$, quotiented by the group of
reflections in these planes. The image of the reduced moment map
$\wt{J}_A:M_{\red} \ra \wt{gl}(2)_-$ is
$$
\eqalignno{
{\cal N}(\l)= {\cal N}_0(\l) =\wt{J}_{A}(F,G)=&
{ \pmatrix{\scst -x_1 y_1 - \mu_1 &\scst -y_1^2 +{\mu_1^2\over x_1^2} \cr
       \scst x_1^2  &\scst  x_1y_1 -\mu_1} \over 2\l} +
{ \pmatrix{\scst -x_2 y_2 - \mu_2 &\scst -y_2^2 +{\mu_2 ^2\over x_2^2} \cr
      \scst  x_2^2  & \scst x_2y_2 -\mu_2} \over 2(\l - 1)}  \cr
 & \quad +
{ \pmatrix{\scst -x_3 y_3 - \mu_3 &\scst y_3^2 - {\mu_3^2 \over x_3^2} \cr
       \scst - x_3^2  & \scst x_3y_3 - \mu_3} \over 2(\l - t)}. &(3.55)}
$$
Choose the Hamiltonian
$$
\eqalignno{
H = H_3 = &{1\over 4 \pi i} \oint_{\l=t} \tr({\cal N}(\l))^2d\l
= {\tr(N_1 N_3) \over t}  + {\tr(N_1 N_3) \over t-1} \cr
= &\quad {1\over 4t}\left[\left(x_1y_3 + x_3 y_1\right)^2 -
\mu_1^2{x_3^2 \over x_1^2}-\mu_3^2{x_1^2 \over x_3^2} + 2\mu_1 \mu_3 \right]
\cr
 & +{1\over 4(t-1)}\left[\left(x_2y_3 + x_3 y_2\right)^2 -
\mu_2^2{x_3^2  \over x_2^2}-\mu_3^2{x_2^2 \over x_3^2} + 2\mu_2 \mu_3 \right] .
&
(3.56) }
$$
 Let
$$
\eqalignno{
a & = {1\over 2} \sum_{i=1}^3 x_iy_i  & (3.57a)\cr
b & = {1\over 2} (y_1^2 +y_2^2 - y_3^2) - {\mu_1^2 \over2x_1^2}-{\mu_2^2\over
2x_2^2}
+{\mu_3^2 \over2 x_3^2}& (3.57b)\cr
c &= {1\over 2} (x_1^2+ x_2^2 - x_3^2). & (3.57c) }
$$
These are the generators of the constant $Sl(2)$ conjugation
 action$$
g: {\cal N}_0(\l) \lra g {\cal N}_0(\l) g^{-1},  \eqno(3.58)
$$
and satisfy
$$
\{a,b\} =b, \quad \{c,a\} =c, \quad \{b,c\} = -2 a.  \eqno(3.59)
$$
(In this case, since $Y=0$, $G_Y = Gl(2)$, but the trace term acts trivially.)
The Hamiltonian (3.56) is invariant under this $Sl(2)$-action
$$
\{a, H\} = \{b, H\} = \{c, H\} = 0, \eqno(3.60)
$$
since the elements of the spectral ring $\IYA$ are $G_Y$ invariant.

  The monodromy preserving deformations  generated by $H$ are then determined
by the
commutativity of the operators ${\cal D}_\l = {\di \over \di \l} - {\cal
N}(\l)$ and
${\cal D}_t$, with ${\cal N}(\l)$ given by eq.~(3.55) and
$$
{\cal D}_t = {\di \over \di t} +
{ \pmatrix{\scst -x_3 y_3 - \mu_3 &\scst y_3^2 - {\mu_3^2 \over x_3^2} \cr
       \scst - x_3^2  & \scst x_3y_3 - \mu_3} \over 2(\l - t)}.  \eqno(3.61)
$$
They also preserve the monodromy of the ``dual'' operator
${\cal D}_z = {\di \over \di z}  - {\cal M}(z)$, where ${\cal M}(z)$ is
determined by
restricting the  moment map $\wt{J}_Y: M_{\red} \ra \wt{gl}(3)_-$ to the
submanifold $M_A
\ss M$
$$
\eqalignno{
{\cal M}(z) = &-A + {\cal M}_0(z) = -A + \wt{J}_Y(F,G)
=-\pmatrix{\scst 0 &\scst 0 &\scst 0 \cr
\scst 0 &\scst 1 &\scst 0 \cr
\scst 0 &\scst 0 &\scst t} \cr
& +{1\over 2z}\pmatrix{
\scst 2\mu_1 &\scst x_1y_2-x_2y_1 + {\mu_1x_2\over x_1} + {\mu_2x_1\over x_2}
&\scst x_1y_3 + x_3y_1 -{\mu_1x_3\over x_1}+{\mu_3x_1\over x_3} \cr
\scst x_2y_1-x_1y_2 + {\mu_1x_2\over x_1} +{\mu_2x_1\over x_2}
 &\scst 2\mu_2 &\scst x_2y_3+ x_3 y_2  -{\mu_2x_3\over x_2}+ {\mu_3x_2\over
x_3} \cr
\scst x_3y_1 + x_1y_3 +{\mu_1x_3\over x_1} -{\mu_3x_1\over x_3}
&\scst x_3y_2 + x_2y_3 +{\mu_2x_3\over x_2}-{\mu_3x_2\over x_3}
&\scst 2\mu_3}.&(3.62)}
$$
 In this $\wt{gl}(3)_-$ representation, the quantities
$$
\eqalignno{
\tr(FG^T)&=\mu_1+\mu_2+\mu_3 &(3.63a)\cr
\tr(FG^T)^2&=\tr(G^TF)^2 = 2(a^2 - bc) +{1\over 2}(\sum_{i=1}^3\mu_i)^2
&(3.63b)\cr
 \det(FG^T) &= 0 &(3.63c)}
$$
are the Casimir invariants, while the individual elements $\mu_1, \mu_2, \mu_3$
are not
Casimirs, but generators of the stabilizer $G_A \ss Gl(3)$ of $A$, and hence
constants of motion.  Thus, what appeared before as Casimirs on $\wt{gl}(2)_-$
become
elements of the spectral ring $\IAY$, while the element of $\IYA$ given by
eq.~(3.63b) becomes a Casimir on $\wt{gl}(3)_-$.

   Viewing $H_3$ now as a function of ${\cal M}$, we have, from eq.~(2.45 )
$$
\eqalign{
&(dH_3)_+ = \pmatrix{\scst 0 &\scst 0 &\scst 0 \cr
\scst 0 &\scst 0 &\scst 0 \cr
\scst 0 &\scst 0 &\scst -z} \cr
& \quad  +{1\over 2}\pmatrix{
\scst 0 & \scst 0
&\scst {1\over t}( x_1y_3 + x_3y_1  -{\mu_1x_3\over x_1} +{\mu_3x_1\over
x_3})\cr
\scst 0 &\scst 0 &\scst{1\over t-1}(x_2y_3+ x_3y_2 -{\mu_2x_3\over x_2}
 + {\mu_3x_2\over x_3} ) \cr
\scst {1\over t}( x_3y_1 + x_1y_3 +{\mu_1x_3\over x_1} -{\mu_3x_1\over x_3})
&\scst {1\over t-1}( x_3y_2 + x_2y_3+{\mu_2x_3\over x_2}-{\mu_3x_2\over x_3})
&\scst 0}.} \eqno(3.64)
$$
The dual monodromy preserving representation is therefore given by the
commutativity of the operators ${\cal D}_z = {\di \over \di z} - {\cal M}(z)$,
and ${\cal
D}^*_t$, with ${\cal M}(z)$ given by eq.~(3.62), and
$$
\eqalign{
{\cal D}_t^* &={\di \over \di t} -(dH_3)_+ + dH_v \cr
& ={\di \over \di t}
+ \pmatrix{\scst 0 &\scst 0 &\scst 0 \cr
\scst 0 &\scst 0 &\scst 0 \cr
\scst 0 &\scst 0 &\scst z}  \cr
& -{1\over 2}\pmatrix{
\scst {1\over t}(\mu_3 -{\mu_1 x_3^2 \over x_1^2} ) & \scst 0
&\scst {1\over t}( x_1y_3 + x_3y_1-{\mu_1x_3\over x_1} +{\mu_3x_1\over x_3}
)\cr
\scst 0 &\scst {1\over t-1}(\mu_3 -{\mu_2 x_3^2 \over x_2^2}) &\scst{1\over
t-1}(x_2y_3+
x_3y_2  -{\mu_2x_3\over x_2} + {\mu_3x_2\over x_3} ) \cr
\scst {1\over t}( x_3y_1 + x_1 y_3 +{\mu_1x_3\over x_1} -{\mu_3x_1\over x_3})
&\scst {1\over t-1}( x_3y_2 + x_2 y_3+{\mu_2x_3\over x_2}-{\mu_3x_2\over x_3})
&\scst\scst \scst{1\over t}(\mu_1-{\mu_3 x_1^2 \over x_3^2} ) +
{1\over t-1}(\mu_2 -{\mu_3 x_2^2 \over x_3^2}) }.}  \eqno(3.65)
$$
Here
$$
dH_v =\pmatrix{\scst {1\over 2t}({\mu_1 x_3^2 \over x_1^2} -\mu_3) &\scst 0
&\scst 0 \cr
         \scst 0 & \scst {1\over 2(t-1)}({\mu_2 x_3^2 \over x_2^2} -\mu_3) &
\scst 0 \cr
         \scst 0 &\scst 0 & \scst{1\over 2t}({\mu_3 x_1^2 \over x_3^2} -\mu_1)
+
{1\over 2(t-1)}({\mu_3 x_2^2 \over x_3^2} -\mu_2)} \eqno(3.66)
$$
is the element of the diagonal subalgebra (corresponding to $G_A$) that must be
added in
order that the lift of the $G_A$-reduced Hamiltonian vector field on $M_{\red}$
be tangential
to $M_A$.

    To obtain the $Sl(2)$-reduced system, first choose the level set
$$
b=c=0, \eqno(3.67)
$$
and again define the ``spectral Darboux coordinates'' $(u,w)$ by
$$
2{\cal N}(\l)_{21} =  \left({x_1^2 \over \l }+ {x_2^2 \over \l -1}
  - {x_3^2 \over \l -t}\right)
 = {w(u-\l) \over a(\l)} \eqno(3.68)
$$
on this level set, where
$$
\eqalignno{
w & = (1+t)x_1^2 + t x_2^2 - x_3^2 & (3.69a) \cr
u &= {t x_1^2\over w}   & (3.69b) \cr
a(\l)& := \l(\l-1)(\l-t). & (3.69c)}
$$
Thus $(u,w)$ are elliptic-hyperbolic coordinates on the cone
$$
x_1^2 + x_2^2 =x_3^2.  \eqno(3.70)
$$
On the invariant manifold defined by $b=c=0$, the symplectic form reduces to
$$
\o_{\red}\vert_{(b,c)=(0,0)}=d(\log w) \wedge da  + du \wedge dv , \eqno(3.71)
$$
where
$$
v= {1\over 2} \left({x_1y_1 -\mu_1 \over u}+{x_2y_2-\mu_2 \over u-1}+
{x_3y_3-\mu_3 \over u-t}\right)\eqno(3.72)
$$
is the momentum coordinate conjugate to $u$. (Note that the slight difference
between this
choice and that of eq.~(3.40) results in a polynomial form for the
Hamiltonian.)

    Restricting to the level set $a=a_0$, we have
$$
\o_{\red}\vert_{(a,b,c)=(a_0,0,0)} = du \wedge dv,
 \eqno(3.73)
$$
which is the reduction of the symplectic form under the
$Sl(2)$-action generated by $(a,b,c)$. The coordinates $(u,v)$  project to the
quotient
under the action of the stability group of the image $(a_0,0,0)$ of this
$sl(2)$ moment
map, since they satisfy $\{u,a\}=\{v,a\}=0$ on  the level set
$(a,b,c)=(a_0,0,0)$.

To compute the Hamiltonian in terms of the reduced coordinates, we write
$$
{1\over 2} \tr({\cal N}(\l))^2)={P_0 + P_1 \l \over a(\l)} +  {\mu_1^2 \over
2\l^2}+{\mu_2^2 \over 2(\l-1)^2}+{\mu_3^2 \over 2(\l-t)^2}. \eqno(3.74)
$$
where
$$
P_1 = a^2 + {1\over 4} \left(\sum_{i=1}^3 \mu_i\right)^2
- {1\over 2}\sum_{i=1}^3\mu_i^2.  \eqno(3.75)
$$
and evaluate
$$
\eqalignno{
{1\over 4 \pi i} \oint_{\l=u}{a(\l)\tr({\cal N}(\l))^2 \over \l-u}d\l
 &= {1\over 2}  u(u-1)(u-t)  {\cal N}^2(u)&(3.76) \cr
&= P_0 +P_1u + {\mu_1^2(u-1)(u-t)\over2 u}+{\mu_2^2u(u-t)\over 2(u-1)}+
{\mu_3^2u(u-1)\over 2(u-t)},}
$$
where the integral is around a circle containing only the pole at $\l=u$. Since
$$
{x_1^2 \over u}+{x_2^2 \over u -1}-{x_3^2 \over u -t}=0, \eqno(3.77)
$$
we have
$$
{1\over 2} \tr{\cal N}^2(u) = v^2 + v \left({\mu_1 \over u} + {\mu_2 \over u
-1}
+{\mu_3 \over u -t}\right) +
{1\over 2}\left({\mu_1 \over u} + {\mu_2 \over u -1}
+{\mu_3 \over u -t}\right)^2,  \eqno(3.78)
$$
and hence,
$$
\eqalign{
P_0 =& u(u-1)(u-t) v^2 +v\left(\mu_1 (u-1)(u-t)+\mu_2 u(u-t)
+\mu_3 u(u-1) \right) \cr
& + \mu_1\mu_2 (u-t) + \mu_2\mu_3 u + \mu_1\mu_3 (u-1)  -P_1u.}  \eqno(3.79)
$$
{}From (3.56), (3.79)
$$
\eqalignno{
H & = {P_0+P_1 t\over t(t-1)} \cr
&={1\over t(t-1)}[u(u-1)(u-t)v^2
  + v\left(\mu_1(u-1)(u-t) + \mu_2u(u-t)
+ \mu_3u(u-1)\right) \cr
&\qquad +\mu_1\mu_2(u-t) +\mu_2\mu_3 u+ \mu_1\mu_3(u-1) + (t -u) P_1]. &(3.80)}
$$
The reduced form of Hamilton's equations are therefore
$$
\eqalignno{
{du\over dt}&= {1\over t(t-1)}\left[2u(u-1)(u-t)v + \mu_1(u-1)(u-t) +
\mu_2u(u-t)
+ \mu_3 u(u-1)\right]  & (3.81a)\cr
{dv\over dt}&= -{1\over t(t-1)}[\left(u(u-1) +u(u-t) +(u-1)(u-t)\right)v^2 \cr
&\quad +v\left(\mu_1(2u-t-1) +\mu_2(2u-t) +\mu_3 (2u-1)\right)
+\mu_1 \mu_2 +\mu_2 \mu_3 +\mu_1 \mu_3 - P_1].\cr &  & (3.81b)}
$$
Upon elimination of $v$, this gives $P_{VI}$:
$$
\eqalignno{
{d^2u\over dt^2}& = {1\over2} \left({1\over u} + {1\over u-1} +{1\over
u-t}\right)\left({du\over dt}\right)^2 - \left({1\over t} + {1\over t-1}
+{1\over
u-t}\right){du\over dt}  \cr
& \qquad +{u(u-1)(u-t) \over t^2 (t-1)^2}
\left(\a +\b{t\over u^2} + \g{t-1\over (u-1)^2} + \d{t(t-1)\over
(u-t)^2}\right), &
(3.82)}
$$
where
$$
\a =2a^2, \quad \b= -{1\over 2}\mu_1^2,
\quad \g= {1\over 2} \mu_2^2,  \quad \d= -{1\over 2}\mu_3^2. \eqno(3.83)
$$
 \bigskip
\noindent {\bf 4. Generalizations} \hfill

   The approach developed in {\bf[AHP, AHH1]} is equally valid for
matrices $A$, $Y$ that are nondiagonalizable, giving rise to isospectral
deformations of
matrices of the form
$$
\eqalignno{
{\cal N}(\l) &= Y + \sum_{i=1}^n \sum_{l_i =1}^{n_i}
{N_{i,p_i} \over(\l -\a_i)^{l_i}}, &(4.1a) \cr
{\cal M}(z) &= -A + \sum_{a=1}^r \sum_{m_a =1}^{r_a}
{M_{a,m_a} \over(z -y_a)^{m_a}}.  &(4.1b)}
$$A straightforward generalization of the moment map construction may also be
made,
yielding the more general forms
$$
\eqalignno{
{\cal N}(\l) &= \sum_{l_{0} =0}^{n_0} Y_{l_0} \l^{l_0}
+ \sum_{i=1}^n \sum_{l_i=1}^{n_i} {N_{i,p_i} \over(\l -\a_i)^{l_i}}  & (4.2a)
\cr
{\cal M}(z) &=  \sum_{m_{0} =0}^{r_0} A_{m_0} z^{m_0}
 + \sum_{a=1}^r \sum_{m_a =1}^{r_a}
{M_{a,m_a} \over(z -y_a)^{m_a}}.  &(4.2b)}
$$
Such ${\cal N}(\l)$, ${\cal M}(z)$  may be viewed as elements, respectively, of
subspaces
$\gYA \ss \wt{gl}(r)^*$, $\gAY \ss \wt{gl}(N)^*$ defined by the rational
structure
appearing in eqs.~(4.2a,b). These are Poisson subspaces with respect to the Lie
Poisson
bracket on $\wt{gl}(r)^*$ (resp. $\wt{gl}(N)^*$) corresponding to the Lie
bracket:
$$
[X,Y]_R :={1\over 2}[RX,Y] + {1\over 2}[X, RY],  \eqno(4.3)
$$
where
$$
R :=P_+ -P_-  \eqno(4.4)
$$
is the {\it classical $R-$matrix} given by the difference of the projection
operators
$P_{\pm}$ to the subalgebras $\wt{gl}(r)_{\pm}$ (resp. $\wt{gl}(N)_{\pm}$). The
$R$-matrix version of the AKS theorem {\bf [S]} again implies that results of
the type (i)
and (ii) (eq.~(2.7)) hold for the autonomous systems generated by elements of
the ring
$\IYA$  (resp $\IAY$)  obtained by restriction of the ring of $\Ad^*$-invariant
polynomials on $\wt{gl}(r)^*$ (resp. $\wt{gl}(N)^*$  to $\gA$ (resp. $\gY$).
Isomonodromic deformations of operators of the type ${\cal D}_\l = {\di \over
\di \l} -
{\cal N}(\l)$, where ${\cal N}(\l)$ is of the form (4.2a),  were the subject of
the series
of papers {\bf [JMU], [JM]}. They are required, in particular, for the
isomonodromy formulation of the remaining Painlev\'e transcendent equations
$P_I$ --
$P_{IV}$ ({\bf [JM]}) and for the Hamiltonian dynamics governing the
level-spacing
distribution functions in random matrix models at the ``edge of the spectrum''
{\bf [TW2]}. A brief discussion of the latter from the loop algebra viewpoint
is given in
{\bf [HTW]}. The Hamiltonian formulation of more general systems of monodromy
preserving  deformations of operators with irregular singular points of
arbitrary order
within the framework of spectral invariants on loop algebras will be addressed
in a
subsequent work.

\bigskip \bigskip
\noindent{\it Acknowledgements.} The author wishes to thank C.~Tracy for
helpful discussions that originally motivated the considerations leading to
the present work.
\bigskip \bigskip
  \centerline{\bf References}

 {\smalltype
\item{\bf [AHH1]} Adams, M.R., Harnad, J. and Hurtubise, J.,
``Dual Moment Maps to Loop Algebras'', {\it Lett.~Math.~Phys.} {\bf 20},
 294-308 (1990).
\item{\bf [AHH2]} Adams, M.R., Harnad, J. and Hurtubise, J.,
`` Darboux Coordinates and Liouville-Arnold Integration
   in Loop Algebras'', CRM preprint (1992),  Commun. Math. Phys. (to appear,
1993).
\item{\bf [AHP]} Adams, M.R., Harnad, J. and Previato, E., ``Isospectral
Hamiltonian Flows in Finite and Infinite Dimensions I. Generalised Moser
Systems
and Moment Maps into Loop Algebras'',  {\it Commun.~Math.~Phys.} {\bf 117},
451-500 (1988).
\item{\bf [HHM]} Harnad, J.,  Hurtubise, J. and Marsden, J.E.,
``Reduction of Hamiltonian Systems with Discrete Symmetries'',
preprint CRM (1992).
\item{\bf [HTW]} Harnad, J., Tracy, C., and Widom, H.,  ``Hamiltonian Structure
of
Equations  Appearing in Random Matrices'',  preprint CRM-1846 (1993),
hep-th/9301051.
\item{\bf [IIKS]} Its, A.~R.~, Izergin, A.~G.~, Korepin, V.~E.~ and Slavnov,
N.~A.,
``Differential Equations for quantum correlation functions'',
Int.~J.~Mod.~Phys. {\bf
B4}, 1003--1037 (1990).
 \item{\bf[JMMS]} Jimbo, M., Miwa, T., M\^ori, Y. and Sato, M.,, ``Density
Matrix of an Impenetrable Bose Gas and the Fifth Painlev\'e Transcendent'',
{\it Physica} {\bf 1D}, 80-158 (1980).
 \item{\bf[JMU]} Jimbo, M., Miwa, T., Ueno, K., ``Monodromy Preserving
Deformation of Linear Ordinary Differential Equations with Rational
Coeefficients I.'', {\it Physica} {\bf 2D}, 306-352 (1981).
 \item{\bf[JM]} Jimbo, M., Miwa, T., ``Monodromy Preserving
Deformation of Linear Ordinary Differential Equations with Rational
Coeefficients II, III.'', {\it Physica} {\bf 2D}, 407-448 (1981); {\it ibid.},
{\bf 4D}, 26-46 (1981).
\item{\bf [M]} Moore, G.~, ``Matrix Models of 2D Gravity and Isomonodromic
Deformation'',
{\it Prog.~Theor.~Phys.} Suppl. No. {\bf 102}, 255-285 (1990).
 \item{\bf[Ok]} Okamoto, K., ``The Painlev\'e equations and the Dynkin
Diagrams'',
in: {\it Painlev\'e Transcendents. Their Asymptotics and Physical
Applications}, NATO ASI
Series B, Vol. {\bf 278}, 299-313 , ed. D.~Levi and P.~Winternitz, Plenum
Press, N.Y.
(1992).
\item{\bf [RS]} Reyman, A.~G., Semenov-Tian-Shansky, A.,
``Current algebras and nonlinear partial differential equations.'',
Soviet.~Math.~Dokl. {\bf 21}, 630-634 (1980); `` A family of Hamiltonian
structures, hierarchy of Hamiltonians, and reduction for first-order matrix
differential
operators'' Funct.~Anal.~Appl. {\bf 14}, 146-148 (1980).
 \item{\bf [S]} Semenov-Tian-Shansky, M.~A., ``What is a Classical R-Matrix'',
{\it Funct.~Anal.~Appl.} {\bf 17}, 259-272 (1983);
``Dressing Transformations and Poisson Group Actions'',
{\it Publ. RIMS Kyoto Univ.} {\bf 21}, 1237-1260 (1985).
\item{\bf [TW1]}Tracy, C., and Widom, H., ``Introduction to Random
Matrices'',   UCD preprint ITD 92/92-10 (1992), to appear in {\it VIIIth
Scheveningen Conf. Proc.},
Springer Lecture Notes in Physics.
\item{\bf [TW2]}Tracy, C., and Widom, H., ``Level Spacing Distribution
Functions
and the Airy Kernel'',   UCD preprint ITD 92/93-9 (1992).
\item{\bf [W]} Weinstein, A., ``The Local Structure of Poisson Manifolds'',
J.~Diff.~Geom. {\bf 18}, 523-557 (1983)
.%
\item {}}
\vfill \eject

\end